\documentclass[11pt,a4paper]{article}
\pdfoutput=1

\usepackage{jheppub}

\usepackage{amsmath}
\usepackage{bbm}
\usepackage{float}
\usepackage{graphicx}	
\usepackage{hyperref}
\usepackage{mathtools}
\usepackage{cancel}

\usepackage[normalem]{ulem}

\usepackage{multirow}
\usepackage{rotating}
\usepackage{subcaption}

\def\lsim{\raise0.3ex\hbox{$\;<$\kern-0.75em\raise-1.1ex
\hbox{$\sim\;$}}}
\def\gsim{\raise0.3ex\hbox{$\;>$\kern-0.75em\raise-1.1ex
\hbox{$\sim\;$}}}

\def\thetitle{ 
Symmetry Finder: A method for hunting symmetry in neutrino oscillation \\
}

\title{\thetitle}
\hypersetup{pdftitle={\thetitle}}
\author{Hisakazu Minakata}

\affiliation{
Center for Neutrino Physics, Department of Physics, Virginia Tech, Blacksburg, Virginia 24061, USA \\
}

\emailAdd{hisakazu.minakata@gmail.com}

\date{\today}

\abstract{Symmetry in neutrino oscillation serves for a better understanding of the physical properties of the phenomenon. We present a systematic way of finding symmetry in neutrino oscillation, which we call {\em Symmetry Finder} (SF). By extending the known framework in vacuum into a matter environment, we derive the SF equation, a powerful machinery for identifying symmetry in the system. After learning lessons on symmetry in the Zaglauer-Schwarzer system with a matter equivalent of the vacuum symmetry, we apply the SF method to the Denton {\it et al.} (DMP) perturbation theory to first order. We show that the method is so powerful that we uncover the eight reparametrization symmetries with the $1 \leftrightarrow 2$ state exchange in DMP, denoted as IA, IB, $\cdot \cdot \cdot$, IVB, all new except for IA. The transformations consist of the both fundamental and dynamical variables, indicating their equal importance. It is also shown that all the symmetries discussed in this paper can be understood as the Hamiltonian symmetries, which ensures their all-order validity and applicability to varying density matter. 

}

\begin{document} 

\maketitle

\section{Introduction}
\label{sec:introduction}

In theory of neutrino oscillation symmetry plays an important role, yielding useful relations between the observables. CPT symmetry implies that the neutrino and antineutrino probabilities are equal, $P(\nu_{\beta} \rightarrow \nu_{\alpha}) = P(\bar{\nu}_{\beta} \rightarrow \bar{\nu}_{\alpha})$ in vacuum. Time reversal T: Even though T symmetry itself is broken in nature~\cite{Christenson:1964fg}, it yields a generalized symmetry, an invariance of the $S$ matrix under T transformation accompanied with complex conjugation of the complex numbers in the theory. The generalized T symmetry holds not only in vacuum but also in matter. As symmetry often brings us useful consequences, finding symmetry and its use in exploration of physical properties of neutrino oscillations should lead us to a deeper understanding of the phenomenon. 

Not so surprisingly, T-odd quantities obey a few remarkable identities and general regularities in the oscillation probability. They include the Naumov~\cite{Naumov:1991ju}  and Toshev~\cite{Toshev:1991ku} identities, whose former implies that the Jarlskog factor in matter must be proportional~\cite{Krastev:1988yu} to the Jarlskog factor in vacuum, $J_r \equiv c_{12} s_{12} c^2_{13} s_{13} c_{23} s_{23}$~\cite{Jarlskog:1985ht}. 
The dependence of the CP phase $\delta$, the lepton counterpart of the quark CP phase~\cite{Kobayashi:1973fv}, on the oscillation probability is also strongly constrained in matter to the single and double harmonics (sine and cosine of $\delta$ and $2 \delta$) \cite{Kimura:2002wd}, which allows simple and informative representations by the CP- or T-conjugate bi-probability diagrams~\cite{Minakata:2001qm,Minakata:2002qe}. For a broader view of CP and T violation see e.g., ref.~\cite{Nunokawa:2007qh}. 

It is the purpose of this paper to present a systematic way of uncovering symmetry in neutrino oscillations. Such a method should be welcome as it makes symmetry consideration handy and thereby contributes to a clearer physics understanding. Now, we must note that symmetry in neutrino oscillation is not necessarily the ones which contain space - time reflection, despite that we have started our description from them. In a perturbative treatment of neutrino oscillation by Denton {\it et al.}~\cite{Denton:2016wmg}, dubbed as DMP hereafter, a relabeling or reparametrization symmetry is found. It is an invariance of the probability under the transformations 
$\lambda_{1} \leftrightarrow \lambda_{2}$, $\cos \psi \rightarrow \mp \sin \psi$, $\sin \psi \rightarrow \pm \cos \psi $. 
Here, $\lambda_{i} / 2E$ ($i=1,2,3$) denote the eigenvalues of the Hamiltonian with $E$ being neutrino energy, and $\psi$ is the mixing angle $\theta_{12}$ in matter. In this paper we will find a new family of these symmetries in the same theory. 

The other examples of the reparametrization symmetry with the eigenstate exchange in the solar- and atmospheric-resonance perturbation theories are later discussed in ref.~\cite{Martinez-Soler:2019nhb}. An interpretation of these symmetries as the ``dynamical symmetry'',\footnote{
A dynamical symmetry is the symmetry that has no hint in the Hamiltonian of the system, but the one which indeed arises after the system is solved. The symmetry often comes with the transformations written by the variables that are used to diagonalize the Hamiltonian. 
}
as opposed to the Hamiltonian symmetry, is also given in ref.~\cite{Martinez-Soler:2019nhb}. An alternative view of these symmetries as due to rephasing invariance of the $S$ matrix is presented in ref.~\cite{Minakata:2020oxb}. In this paper, we will shed more light on the nature of these relabelling or reparametrization symmetries by bringing the new symmetries in DMP into this context. Through a renewed discussion on symmetry from the Hamiltonian point of view, we will be able to sharpen up our previous statement on the Hamiltonian symmetry. 

We start our discussion from what we call the ``Symmetry Finder'' (SF) equation in vacuum~\cite{Parke:2018shx}. It is nothing but the expression of the flavor basis state (i.e., wave function) $\nu$ in terms of the mass eigenstate $\check{\nu}$ in the following two different ways, 
\begin{eqnarray} 
&&
\nu = U (\theta_{23}, \theta_{13}, \theta_{12}, \delta) \check{\nu} 
= U (\theta_{23}^{\prime}, \theta_{13}^{\prime}, \theta_{12}^{\prime}, \delta^{\prime}) \check{\nu}^{\prime}, 
\label{SF-eq-vacuum}
\end{eqnarray}
where the quantities with ``prime'' imply the transformed ones, and $\check{\nu}^{\prime}$ may involve eigenstate exchanges and/or rephasing of the wave functions. Since the SF equation represents the same flavor state by the two different sets of the physical parameters, it implies a symmetry. See section~\ref{sec:symmetry-finder-vacuum} for more details. 

We utilize the SF equation in its generalized form in matter to uncover new symmetries associated with the mass-eigenstate exchange $1 \leftrightarrow 2$. We first treat the exact solution of neutrino oscillation under a uniform matter density, the Zaglauer-Schwarzer system~\cite{Zaglauer:1988gz} in section~\ref{sec:symmetry-finder-matter}. Then, as the highlight of SF symmetry discussion, it will be applied to the DMP framework~\cite{Denton:2016wmg} in sections~\ref{sec:symmetry-finder-DMP} and \ref{sec:hamiltonian-view}. The DMP perturbation theory gives an approximate but very accurate description of neutrino oscillation in matter with uniform matter density~\cite{Parke:2019vbs}, and at the same time displays a clear physical picture which covers the entire kinematical region of the terrestrial neutrino experiments~\cite{Minakata:2020oxb}. In fact, we will uncover the eight symmetries whose seven are new, with interesting varying features of the mixed fundamental and dynamical symmetries. See the summary Table~\ref{tab:DMP-symmetry} in section~\ref{sec:eight-symmetries}.  
Due to the frequent usage of the term such as ``the DMP system'' or ``the DMP framework'', we sometimes abbreviate them as just ``DMP'' hereafter. 

A few words on the type of symmetry we are going to discuss in this paper. It does not involve the transformations that connect our world to the other one, such as $\Delta m^2_{31} \rightarrow - \Delta m^2_{31}$, or $\theta_{12} \leq \frac{\pi}{4}$ to $\theta_{12} \geq \frac{\pi}{4}$.\footnote{
In contrast, it is argued in ref.~\cite{Minakata:2010zn} that the sign-$\Delta m^2$ and the $\theta_{23}$ octant degeneracies can be understood as a consequence of the symmetries, though approximate ones, which connect the systems with differing signs of $\Delta m^2_{31}$~\cite{Minakata:2001qm}, and the worlds with $\theta_{23} \leq \frac{\pi}{4}$ and $\theta_{23} \geq \frac{\pi}{4}$~\cite{Fogli:1996pv}. 
}
By restricting the symmetry transformation into the ones which keeps the system to remain in our world it inevitably takes the form of reparametrization or relabelling symmetry. Implications of the newly observed reparametrization symmetry are discussed in sections~\ref{sec:hamiltonian-view} and \ref{sec:nature-symmetry-matter}. 

\section{The three neutrino evolution in matter in the $\nu$SM}
\label{sec:3nu-SM}

In this paper, we restrict ourselves into discussions within the neutrino-mass-embedded Standard Model, $\nu$SM for short. We discuss symmetry which exists in the oscillation probability in the standard three-flavor neutrino system defined by the Hamiltonian in the flavor basis 
\begin{eqnarray}
H= 
\frac{ 1 }{ 2E } \left\{ 
U \left[
\begin{array}{ccc}
m^2_{1} & 0 & 0 \\
0 & m^2_{2} & 0 \\
0 & 0 & m^2_{3}
\end{array}
\right] U^{\dagger}
+
\left[
\begin{array}{ccc}
a(x) & 0 & 0 \\
0 & 0 & 0 \\
0 & 0 & 0
\end{array}
\right] 
\right\}, 
\label{flavor-basis-hamiltonian}
\end{eqnarray}
where $E$ is neutrino energy, and $m_{i}$ (i=1,2,3) the neutrino masses of $i$-th mass eigenstate. 
In eq.~\eqref{flavor-basis-hamiltonian}, $U \equiv U_{\text{\tiny MNS}}$ denotes the standard $3 \times 3$ lepton flavor mixing matrix~\cite{Maki:1962mu} which relates the flavor neutrino states to the vacuum mass eigenstates as $\nu_{\alpha} = U_{\alpha i} \nu_{i}$, where $\alpha$ runs over $e, \mu, \tau$, and the mass eigenstate index $i$ runs over $1,2,$ and $3$. 
The functions $a(x)$ in \eqref{flavor-basis-hamiltonian} denote the Wolfenstein matter potential \cite{Wolfenstein:1977ue} due to charged current (CC) reactions 
\begin{eqnarray} 
a (x) &=&  
2 \sqrt{2} G_F N_e E \approx 1.52 \times 10^{-4} \left( \frac{Y_e \rho (x) }{\rm g\,cm^{-3}} \right) \left( \frac{E}{\rm GeV} \right) {\rm eV}^2.  
\label{matt-potential}
\end{eqnarray}
Here, $G_F$ is the Fermi constant, $N_e$ is the electron number density in matter. $\rho (x)$ and $Y_e$ denote, respectively, the matter density and number of electron per nucleon in matter. 

In sections~\ref{sec:symmetry-finder-matter} and \ref{sec:symmetry-finder-DMP}, our treatment assumes the uniform matter density approximation. Whereas in section~\ref{sec:hamiltonian-view} where we discuss the Hamiltonian view of the symmetries, we get rid of this restriction but keeping variation of matter density sufficiently mild such that the adiabatic approximation applies. 

\subsection{The SOL convention of the flavor mixing matrix}
\label{sec:SOL}

We use, throughout this paper, the SOL convention~\cite{Parke:2018shx,Martinez-Soler:2018lcy} for the lepton mixing matrix 
\begin{eqnarray} 
&& 
U_{\text{\tiny SOL}} 
= 
\left[
\begin{array}{ccc}
1 & 0 &  0  \\
0 & c_{23} & s_{23} \\
0 & - s_{23} & c_{23} \\
\end{array}
\right] 
\left[
\begin{array}{ccc}
c_{13}  & 0 & s_{13} \\
0 & 1 & 0 \\
- s_{13} & 0 & c_{13} \\
\end{array}
\right] 
\left[
\begin{array}{ccc}
c_{12} & s_{12} e^{ i \delta} & 0  \\
- s_{12} e^{- i \delta} & c_{12} & 0 \\
0 & 0 & 1 \\
\end{array}
\right] 
\nonumber \\
&\equiv& 
U_{23} (\theta_{23}) U_{13} (\theta_{13}) U_{12} (\theta_{12}, \delta ), 
\label{U-MNS-SOL}
\end{eqnarray}
in which $e^{ \pm i \delta}$ is attached to sine of the ``solar angle'' $\theta_{12}$, with $\delta$ being the lepton CP phase. We note that $s_{ij} \equiv \sin \theta_{ij}$ etc. are the common abbreviated notations. We will see in section~\ref{sec:symmetry-finder-vacuum} that the SOL convention is the most convenient one to discuss the symmetries which involve the 1-2 mass eigenstate exchange and the CP phase $\delta$. With the SOL convention, we obtain the exactly the same oscillation probability as the one calculated by using the well known Particle Data Group (PDG)~\cite{Zyla:2020zbs} convention, as to be seen immediately below. Thus, there is no reason for being afraid of using the SOL convention, despite its unfamiliar status.\footnote{
Yet, the SOL convention of the mixing matrix has been used in an analysis of the parameter correlations in the theory with unitarity violation~\cite{Martinez-Soler:2019noy}. 
}

The mixing matrix in the SOL and the PDG conventions are related by~\cite{Martinez-Soler:2018lcy}
\begin{eqnarray}
U_{\text{\tiny SOL}} 
&=& 
\left[
\begin{array}{ccc}
1 & 0 &  0  \\
0 & e^{ - i \delta} & 0 \\
0 & 0 & e^{ - i \delta} \\
\end{array}
\right] 
U_{\text{\tiny PDG}} 
\left[
\begin{array}{ccc}
1 & 0 &  0  \\
0 & e^{ i \delta} & 0 \\
0 & 0 & e^{ i \delta} \\
\end{array}
\right].
\label{U-SOL-PDG}
\end{eqnarray}
Then, their flavor-basis Hamiltonian and the neutrino states in the both conventions are related with each other by
\begin{eqnarray}
H_{\text{\tiny SOL}} 
&=&
\left[
\begin{array}{ccc}
1 & 0 &  0  \\
0 & e^{ - i \delta} & 0 \\
0 & 0 & e^{ - i \delta} \\
\end{array}
\right] 
H_{\text{\tiny PDG}} 
\left[
\begin{array}{ccc}
1 & 0 &  0  \\
0 & e^{ i \delta} & 0 \\
0 & 0 & e^{ i \delta} \\
\end{array}
\right], 
\hspace{8mm}
\nu_{\text{\tiny SOL}} 
= 
\left[
\begin{array}{ccc}
1 & 0 &  0  \\
0 & e^{ - i \delta} & 0 \\
0 & 0 & e^{ - i \delta} \\
\end{array}
\right] 
\nu_{\text{\tiny PDG}}, 
\label{H-SOL-PDG}
\end{eqnarray}
where $H_{\text{\tiny SOL}}$ and $\nu_{\text{\tiny SOL}}$
($H_{\text{\tiny PDG}}$ and $\nu_{\text{\tiny PDG}}$) denote the Hamiltonian written with use of the $U$ matrix of the SOL (PDG) convention and the neutrino flavor state under the convention. The $S$ matrix obeys the same relation as that of the Hamiltonian in eq.~\eqref{H-SOL-PDG}. It means that the both Hamiltonian and $S$ matrix are identical up to the phase redefinition of neutrino wave functions. Therefore, the oscillation probability calculated with $H_{\text{\tiny SOL}}$ is identical with the one obtained with the usual PDG convention $U$ matrix, as stated above. 

In section~\ref{sec:symmetry-finder-DMP} we will face with another convention change, the ATM to the SOL conventions of the $U$ matrix, where $e^{ \pm i \delta}$ is attached to $s_{23}$ in the ATM convention. Therefore, we give here the corresponding transformation rule between the ATM and SOL conventions~\cite{Martinez-Soler:2018lcy}: 
\begin{eqnarray}
H_{\text{\tiny SOL}} 
&=&
\left[
\begin{array}{ccc}
1 & 0 &  0  \\
0 & e^{ - i \delta} & 0 \\
0 & 0 & 1 \\
\end{array}
\right] 
H_{\text{\tiny ATM}} 
\left[
\begin{array}{ccc}
1 & 0 &  0  \\
0 & e^{ i \delta} & 0 \\
0 & 0 & 1 \\
\end{array}
\right], 
\hspace{8mm}
\nu_{\text{\tiny SOL}} 
= 
\left[
\begin{array}{ccc}
1 & 0 &  0  \\
0 & e^{ - i \delta} & 0 \\
0 & 0 & 1 \\
\end{array}
\right] 
\nu_{\text{\tiny ATM}}. 
\label{H-SOL-ATM}
\end{eqnarray}
The ATM convention of the $U$ matrix is used in refs.~\cite{Minakata:2015gra,Denton:2016wmg,Martinez-Soler:2018lcy,Minakata:2020oxb}, and the references cited therein. 

\section{Symmetry Finder for neutrino oscillation in vacuum}
\label{sec:symmetry-finder-vacuum}

\subsection{Symmetry Finder equation}
\label{sec:sym-finder-eq}

In vacuum the flavor eigenstate $\nu$ is related to the mass eigenstate $\hat{\nu}$ as  
\begin{eqnarray} 
&& 
\nu 
= 
U_{23} (\theta_{23}) U_{13} (\theta_{13}) 
U_{12} (\theta_{12}, \delta) \hat{\nu}. 
\label{flavor-mass-vacuum}
\end{eqnarray}
Using the SOL convention $U$ matrix in eq.~\eqref{U-MNS-SOL}, one can easily prove the relation~\cite{Parke:2018shx}
\begin{eqnarray} 
&& 
U_{12} ( \theta_{12}, \delta ) 
\left[
\begin{array}{c}
\nu_{1} \\
\nu_{2} \\
\nu_{3} \\
\end{array}
\right] 
= U_{12} \left( \theta_{12} + \frac{\pi}{2}, \delta \right) 
\left[
\begin{array}{c}
- e^{ i \delta } \nu_{2}  \\
e^{ - i \delta } \nu_{1} \\
\nu_{3} \\
\end{array}
\right] 
= 
U_{12} \left( \frac{\pi}{2} - \theta_{12}, \delta \pm \pi \right) 
\left[
\begin{array}{c}
e^{ i \delta } \nu_{2}  \\
- e^{ - i \delta } \nu_{1} \\
\nu_{3} \\
\end{array}
\right].
\nonumber \\
\label{symmetry-finder-vacuum}
\end{eqnarray}
We show below that this equation is a powerful tool for uncovering symmetry in neutrino oscillation in vacuum. For this reason, we denote eq.~\eqref{symmetry-finder-vacuum} and its extension in matter as the {\em ``Symmetry Finder''} (SF) equation. 

It is easy to observe that the relation~\eqref{symmetry-finder-vacuum} implies symmetry~\cite{Parke:2018shx}. The first equality means that use of $\theta_{12}^{\prime} = \theta_{12} + \frac{\pi}{2}$ and the exchanged (and rephased) mass eigenstates $1 \leftrightarrow 2$ produces the same oscillation probability. Since rephasing does not affect the observables, the first equality in eq.~\eqref{symmetry-finder-vacuum} implies the $1 \leftrightarrow 2$ exchange symmetry under the transformation 
\begin{eqnarray} 
&& 
\text{Symmetry IA-vacuum:}
\hspace{6mm}
m^2_{1} \leftrightarrow m^2_{2}, 
\hspace{8mm}
c_{12} \rightarrow - s_{12},
\hspace{8mm}
s_{12} \rightarrow c_{12},
\label{Symmetry-IA-vacuum}
\end{eqnarray} 
where existence of an alternative choice, $c_{12} \rightarrow s_{12}$ and $s_{12} \rightarrow - c_{12}$ ($\theta_{12} \rightarrow \theta_{12} - \frac{\pi}{2}$) is understood. Similarly, the second equality in ~\eqref{symmetry-finder-vacuum} implies the symmetry of the probability under the transformation 
\begin{eqnarray} 
&&
\text{Symmetry IB-vacuum:}
\hspace{6mm}
m^2_{1} \leftrightarrow m^2_{2}, 
\hspace{8mm}
c_{12} \leftrightarrow s_{12}, 
\hspace{8mm}
\delta \rightarrow \delta \pm \pi.
\label{Symmetry-IB-vacuum}
\end{eqnarray} 

Throughout this paper we will observe pairing of the symmetries, the types ``A'' and ``B'', where A does not contain $\delta$, while B does. The question of how the pair should be chosen in the presence of many symmetries are present will be answered in the treatment of the DMP perturbation theory in sections~\ref{sec:symmetry-finder-DMP} and \ref{sec:hamiltonian-view}, in which we will observe eight symmetries. 

There are many ways to confirm that the transformations of Symmetry IA and IB above leave the oscillation probability invariant. One can do it by using the explicit expressions of the probability in vacuum. One can also show the invariance of the $S$ matrix elements~\cite{Parke:2018shx}, or the invariance of the vacuum part of the Hamiltonian~\eqref{flavor-basis-hamiltonian}.\footnote{
In fact, Symmetry IA-vacuum leaves the whole Hamiltonian $H$ in eq.~\eqref{flavor-basis-hamiltonian} invariant, as the matter potential term does not transform. Using this feature the author of ref.~\cite{Zhou:2016luk} discussed how one can organize the perturbative expansion in such a way that the symmetry is respected in each order. }
See the comment at the end of section~\ref{sec:H-RHS-transf}. 
The last path, finding a symmetry by the SF equation followed by confirmation of its validity by examining the Hamiltonian, anticipates the route we will take in our treatment of the symmetries in DMP.  

\subsection{Clarification and interpretation of Symmetry IA and IB in vacuum}
\label{sec:significance}

Now a clarifying remark must be made: 
The notation $\theta_{12}^{\prime} = \theta_{12} + \frac{\pi}{2}$, which is used for simplicity of expression of eq.~\eqref{symmetry-finder-vacuum}, may be confusing because it requires to expand the region of definition of $\theta_{12}$ outside of what is usually taken, $0 \leq \theta_{12} \leq \frac{\pi}{2}$~\cite{Zyla:2020zbs}. While it is possible, we do not choose this option in this paper. What we have meant for Symmetry IA is, therefore, an invariance under the transformation in which $c_{12}$ is replaced by $- s_{12}$, and $s_{12}$ is replaced by $c_{12}$ simultaneously with the $1 \leftrightarrow 2$ state exchange $m^2_{1} \leftrightarrow m^2_{2}$ under the condition that the both initial and transformed $\theta_{12}$ remain into the region $0 \leq \theta_{12} \leq \frac{\pi}{2}$. We apply this principle to all the mass eigenstate exchange symmetries $1 \leftrightarrow 2$ in this paper. 
As a matter of fact, we restrict ourselves to the region of $\theta_{12}$ as $0 \leq \theta_{12} \leq \frac{\pi}{4}$, the favored region~\cite{Mikheev:1986gs} first observed experimentally by Davis {\it et al.}~\cite{Cleveland:1998nv}. Under this prescription, the Symmetry IB in eq.~\eqref{Symmetry-IB-vacuum} has nothing to do with the so called ``dark-side'' discussion~\cite{deGouvea:2000pqg,Fogli:2001wi}. 

Then the question would be: What is the interpretation of Symmetry IA and IB in vacuum? The answer is that the SF equation~\eqref{symmetry-finder-vacuum} offers three different descriptions of the unique (i.e., identical) world. If one computes the flavor basis neutrino wave functions by using eq.~\eqref{flavor-mass-vacuum} with the three different ways of the 1-2 space rotation, one obtains the identical expressions of $\nu_{e}$, $\nu_{\mu}$, and $\nu_{\tau}$ in terms of the three mass eigenstate components $\nu_{1}$, $\nu_{2}$, and $\nu_{3}$. Therefore, Symmetry IA and IB are both the relabeling or reparametrization symmetries $1 \leftrightarrow 2$ associated with the transformations of $\theta_{12}$ and possibly $\delta$. It is intriguing to see whether the similar pairing structure of types A and B prevails in matter. 

\section{Symmetry Finder for neutrino oscillation in matter}
\label{sec:symmetry-finder-matter}

In this and the following two sections, we extend the symmetry finding framework in neutrino oscillation, ``Symmetry Finder'', into the matter environments. 
In neutrino oscillation in matter, we often encounter the situation that all or some of the mixing angles and the CP phase $\delta$ becomes matter-dressed. This is the case in the systems we will treat, all become matter-dressed in the Zaglauer-Schwarzer (ZS) system, and only $\theta_{12}$ and $\theta_{13}$ are altered to matter-affected ones in DMP. For notational simplicity and unity we denote the changes in notations for the mixing parameters from vacuum to matter-dressed ones as 
\begin{eqnarray} 
&&
\theta_{12} \rightarrow \psi, ~~~~~~
\theta_{13} \rightarrow \phi, ~~~~~~
\theta_{23} \rightarrow \eta, ~~~~~~
\delta \rightarrow \tilde{\delta}, 
\label{matter-notation}
\end{eqnarray}
throughout sections~\ref{sec:symmetry-finder-matter}, \ref{sec:symmetry-finder-DMP}, and \ref{sec:hamiltonian-view}, except that $\theta_{23}$ and $\delta$ are kept matter-undressed in the latter two sections. Hereafter, we use the abbreviated notations $c_{\psi} \equiv \cos \psi$, $s_{\phi} \equiv \sin \phi$, and etc. 

\subsection{The Zaglauer-Schwarzer system}
\label{sec:ZS} 

Now we apply the SF method to the ZS system in this section as a simple extension of the SF equation in vacuum. 
The authors of ref.~\cite{Zaglauer:1988gz} exactly solved the three-flavor neutrino evolution in matter with uniform density. The flavor neutrino eigenstate is related to the matter propagation eigenstate $\hat{\nu}$ in the SOL convention as 
\begin{eqnarray} 
&& 
\nu 
= 
U_{23} (\eta) U_{13} (\phi) 
U_{12} (\psi, \tilde{\delta}) \hat{\nu}.
\label{flavor-mass-matter}
\end{eqnarray}
Then, the expressions of the oscillation probability in various channels in matter can be obtained by doing replacement given in eq.~\eqref{matter-notation} in the formulas in vacuum. The expressions of the matter-dressed mixing angles and the CP phase are given in ref.~\cite{Zaglauer:1988gz}. Though they are given in the PDG convention, one can easily translate them into the ones in the SOL convention by using eq.~\eqref{U-SOL-PDG}.

The exact expressions of the three eigenvalues of $2EH$ are obtained in ref.~\cite{Barger:1980tf}, which are denoted here as $\lambda_{i}$ ($i=1,2,3$). Our notation is such that, in the normal mass ordering, $\lambda_{3} > \lambda_{2} > \lambda_{1}$ in region of a large positive matter potential defined in eq.~\eqref{matt-potential}. This prescription for defining $\lambda_{i}$ will be used also in discussion of the DMP system in section~\ref{sec:symmetry-finder-DMP}. 
Thanks to the complete parallelism between neutrino oscillations in vacuum and in matter we can immediately conclude, by using the SF equation in matter with replacement~\eqref{matter-notation} in eq.~\eqref{symmetry-finder-vacuum}, that the matter versions of Symmetry IA and Symmetry IB in the ZS system exist: 
\begin{eqnarray} 
&& 
\text{Symmetry IA-ZS:}
\hspace{6mm}
\lambda_{1} \leftrightarrow \lambda_{2}, 
\hspace{8mm}
c_{\psi} \rightarrow - s_{\psi},
\hspace{8mm}
s_{\psi} \rightarrow c_{\psi}, 
\nonumber \\
&&
\text{Symmetry IB-ZS:}
\hspace{6mm}
\lambda_{1} \leftrightarrow \lambda_{2}, 
\hspace{8mm}
c_{\psi} \leftrightarrow s_{\psi}, 
\hspace{11mm}
\tilde{\delta} \rightarrow \tilde{\delta} \pm \pi.
\label{Symmetry-IA-IB-ZS}
\end{eqnarray} 
Notice that these are the symmetries whose transformations consist only of the matter-dressed variables. None of the fundamental parameters, the ones in the original Hamiltonian in eq.~\eqref{flavor-basis-hamiltonian}, transform. 

\subsection{Fundamental vs. dynamical symmetries} 
\label{sec:terminology}

A clarifying remark is in order on our terminology: In this paper, we often use the terms, ``fundamental variables'' and ``dynamical variables''. The fundamental variables are meant to be the mixing parameters in the flavor basis Hamiltonian in eq.~\eqref{flavor-basis-hamiltonian}. In contrast, dynamical variables are characterized as the variables by which the Hamiltonian is diagonalized exactly (ZS case), or approximately (DMP case)~\cite{Minakata:2020oxb}. In the systems discussed in this paper, therefore, the dynamical variables are the matter-dressed variables. 

A dynamical symmetry is the symmetry without its trace in the Hamiltonian of the system, but the one which indeed arises after the system is solved. Hence, it often happens that the dynamical symmetry has transformations written by the variables that are used to diagonalize the Hamiltonian. On the contrary, if the symmetry is described solely by the fundamental variables, it is called as the ``fundamental symmetry''. We will see later that most of the symmetries in DMP have the mixed fundamental - dynamical character. 

\subsection{Generic $i \leftrightarrow j$ state exchange symmetry?}
\label{sec:ij-exchange}

A natural question would be: Why should we restrict ourselves to the $1 \leftrightarrow 2$ exchange symmetry? How about a $1 \leftrightarrow 3$ exchange symmetry, for example? It is a very good point, but we do not know a complete answer about feasibility of such an extension. 

Suppose that we take a different parametrization of the $U$ matrix such as $U=U_{23} (\theta_{23}^{\prime}) U_{12} (\theta_{12}^{\prime}) U_{13} (\theta_{13}^{\prime}, \delta^{\prime} )$ by exchanging the order of $U_{13}$ and $U_{12}$ rotations, see e.g., ref.~\cite{Denton:2020igp}. Then, the similar treatment can go through in vacuum and in matter, which would lead to an analogous discussion of the $1 \leftrightarrow 3$ exchange symmetry. But, $\theta_{ij}^{\prime}$ ($i,j=1,2,3$) and $\delta^{\prime}$ in this new parametrization is completely different from those in the our conventional parametrization. Rewriting the newly obtained $1 \leftrightarrow 3$ exchange symmetry transformations written with $\theta_{ij}^{\prime}$ ($i,j=1,2,3$) and $\delta^{\prime}$ by the three angles and CP phase in our SOL or PDG convention $U$ matrix would be a formidable task. 

Therefore, in principle, one could discuss the $1 \leftrightarrow 3$ or $2 \leftrightarrow 3$ exchange symmetries in the same way as we do, but no useful output is expected if we write the symmetry transformations by the angles and the CP phase in the $U$ matrix of our usual conventions.  For this reason we restrict our discussions to the $1 \leftrightarrow 2$ exchange symmetry in this paper. 

\section{Symmetry Finder in the DMP system}
\label{sec:symmetry-finder-DMP}

We now discuss the Symmetry Finder (SF) equation and its application to the $1 \leftrightarrow 2$ eigenstate exchange symmetry in the Denton {\it et al.} framework~\cite{Denton:2016wmg}. Our treatment in this section is valid to first order in the DMP perturbation theory. Later in section~\ref{sec:hamiltonian-view}, we discuss Hamiltonian proof of the symmetry, which in fact guarantee its validity to all-orders. 

After brief recollections of the DMP perturbation theory in sections~\ref{sec:DMP-0th} and \ref{sec:DMP-1st}, we introduce the SF equation in DMP in section~\ref{sec:SF-eq-DMP} by taking into account of possible rephasing of both the neutrino flavor and energy eigenstates. By using the SF equation we will find the eight symmetries in DMP, which we call Symmetry I-, II-, III-, and IV-DMP doubled with the pairing of ``A'' (no $\delta$) and ``B'' (with $\delta$) types. All eight symmetries are tabulated in the summary Table~\ref{tab:DMP-symmetry} in section~\ref{sec:eight-symmetries}. 
All of them are the reparametrization symmetries with the $1 \leftrightarrow 2$ state exchange. Invariance of the oscillation probability under the transformations listed in Table~\ref{tab:DMP-symmetry} is verified explicitly by using the probability formulas given in refs.~\cite{Denton:2016wmg,Minakata:2020oxb}, which is however left for the readers as a simple exercise.\footnote{
We recommend arXiv version 3 of ref.~\cite{Minakata:2020oxb} for more explicitly written formulas for verification. }

\subsection{DMP at the zeroth order} 
\label{sec:DMP-0th}

At the leading (zeroth) order in the DMP perturbation theory, the flavor neutrino eigenstate is related to the propagation eigenstate in matter as~\cite{Denton:2016wmg}
\begin{eqnarray} 
&& 
\nu 
= 
U_{23} (\theta_{23}) U_{13} (\phi) 
U_{12} (\psi, \delta) \hat{\nu}.
\label{flavor-mass-DMP-0th}
\end{eqnarray} 
which is identical with the vacuum form~\eqref{flavor-mass-vacuum} apart from replacements $\theta_{12} \rightarrow \psi$ and $\theta_{13} \rightarrow \phi$. Or, starting from eq.~\eqref{flavor-mass-matter} in the ZS system, $\eta$ and $\tilde{\delta}$ take the vacuum values $\theta_{23}$ and $\delta$. It must be a good approximation, given that the matter effect modification is modest for them~\cite{Zaglauer:1988gz,Blennow:2013rca}. Then, at the leading order, the DMP system possesses the symmetries akin to Symmetry IA-ZS and IB-ZS in eq.~\eqref{Symmetry-IA-IB-ZS}, but with $\eta$ and $\tilde{\delta}$ replaced by $\theta_{23}$ and $\delta$, respectively. 

\subsection{DMP expansion to first order} 
\label{sec:DMP-1st}

However, when the first order correction is added the structure of the SF equation changes. It can be obtained most easily by the so called $V$ matrix method~\cite{Minakata:1998bf} as done in ref.~\cite{Denton:2016wmg}, which takes the form in the SOL convention as\footnote{
In the original reference~\cite{Denton:2016wmg}, the expression of this SF equation~\eqref{SF-eq-DMP-1st} is given in the ATM convention defined in section~\ref{sec:SOL}. The expression in eq.~\eqref{SF-eq-DMP-1st} with~\eqref{mathcal-W} in the SOL convention can be obtained by the transformation in eq.~\eqref{H-SOL-ATM} in section~\ref{sec:SOL}.
}
\begin{eqnarray} 
&& 
\left[
\begin{array}{c}
\nu_{e} \\
\nu_{\mu} \\
\nu_{\tau} \\
\end{array}
\right] 
= 
U_{23} (\theta_{23}) U_{13} (\phi) U_{12} (\psi, \delta) 
\biggl\{
1 + 
\epsilon c_{12} s_{12} \sin ( \phi - \theta_{13} ) 
\mathcal{W} ( \psi, \delta; \lambda_{1}, \lambda_{2} ) 
\biggr\} 
\left[
\begin{array}{c}
\nu_{1} \\
\nu_{2} \\
\nu_{3} \\
\end{array}
\right], 
\nonumber \\
\label{SF-eq-DMP-1st}
\end{eqnarray}
where $\mathcal{W} ( \psi, \delta; \lambda_{1}, \lambda_{2} )$ is defined by 
\begin{eqnarray} 
&&
\mathcal{W} ( \psi, \delta; \lambda_{1}, \lambda_{2} )
\equiv 
\left[
\begin{array}{ccc}
0 & 0 & - s_\psi 
\frac{ \Delta m^2_{ \text{ren} } }{ \lambda_{3} - \lambda_{1} } \\
0 & 0 & c_\psi e^{ - i \delta} 
\frac{ \Delta m^2_{ \text{ren} } }{ \lambda_{3} - \lambda_{2} } \\
s_\psi 
\frac{ \Delta m^2_{ \text{ren} } }{ \lambda_{3} - \lambda_{1} } & 
- c_\psi e^{ i \delta} 
\frac{ \Delta m^2_{ \text{ren} } }{ \lambda_{3} - \lambda_{2} } & 0 \\
\end{array}
\right]. 
\label{mathcal-W}
\end{eqnarray}
In eqs.~\eqref{SF-eq-DMP-1st} and \eqref{mathcal-W}, the notations $\phi$ and $\psi$ are defined in eq.~\eqref{matter-notation}, and their expressions are given in ref.~\cite{Denton:2016wmg}. $\epsilon$ is the unique expansion parameter in the DMP perturbation theory and is defined as 
\begin{eqnarray} 
&&
\epsilon \equiv \frac{ \Delta m^2_{21} }{ \Delta m^2_{ \text{ren} } }, 
\hspace{10mm}
\Delta m^2_{ \text{ren} } \equiv \Delta m^2_{31} - s^2_{12} \Delta m^2_{21},
\label{epsilon-Dm2-ren-def}
\end{eqnarray}
where $\Delta m^2_{ \text{ren} }$ is the ``renormalized'' atmospheric $\Delta m^2$ introduced in ref.~\cite{Minakata:2015gra}. With this added structure that comes from first-order corrections, we need a reformulation of our symmetry hunting scheme using the SF equation. 

\subsection{Symmetry Finder (SF) equation in DMP} 
\label{sec:SF-eq-DMP}

In looking for the solution to the SF equation~\eqref{SF-eq-DMP-1st} we introduce the ansatz 
\begin{eqnarray} 
&& 
\hspace{-6mm}
F \left[
\begin{array}{c}
\nu_{e} \\
\nu_{\mu} \\
\nu_{\tau} \\
\end{array}
\right] 
= 
F 
U_{23} (\theta_{23}) U_{13} (\phi) U_{12} (\psi, \delta) 
%
G^{\dagger} G 
\biggl\{
1 + 
\epsilon c_{12} s_{12} \sin ( \phi - \theta_{13} ) 
\mathcal{W} ( \psi, \delta; \lambda_{1}, \lambda_{2} ) 
\biggr\} 
G^{\dagger} G 
\left[
\begin{array}{c}
\nu_{1} \\
\nu_{2} \\
\nu_{3} \\
\end{array}
\right], 
\nonumber \\
\label{SF-eq-DMP}
\end{eqnarray}
for an alternative expression of the state, analogously as in eq.~\eqref{symmetry-finder-vacuum}. In eq.~\eqref{SF-eq-DMP} we have introduced the flavor-state rephasing matrix $F$, which is defined by 
\begin{eqnarray} 
&&
F \equiv 
\left[
\begin{array}{ccc}
e^{ i \tau } & 0 & 0 \\
0 & e^{ i \sigma } & 0 \\
0 & 0 & 1 \\
\end{array}
\right],
\label{F-def}
\end{eqnarray}
and the generalized $1 \leftrightarrow 2$ state exchange matrix $G$
\begin{eqnarray} 
&& 
G \equiv 
\left[
\begin{array}{ccc}
0 & - e^{ i ( \delta + \alpha) } & 0 \\
e^{ - i ( \delta + \beta) } & 0 & 0 \\
0 & 0 & 1 \\
\end{array}
\right], 
\hspace{8mm}
G^{\dagger} \equiv 
\left[
\begin{array}{ccc}
0 & e^{ i ( \delta + \beta) } & 0 \\
- e^{ - i ( \delta + \alpha) } & 0 & 0 \\
0 & 0 & 1 \\
\end{array}
\right], 
\label{G-def}
\end{eqnarray}
where $\tau$, $\sigma$, $\alpha$, and $\beta$ denote the arbitrary phases. Notice that the rephasing matrices, the both $F$ and $G$ in eqs.~\eqref{F-def} and \eqref{G-def} takes the nonvanishing, nontrivial (not unity) elements in 1-2 sub-sector. It is because we restrict ourselves into the $1 \leftrightarrow 2$ state exchange symmetry. 

Now, we call the readers' attention that since we have introduced the flavor-state rephasing matrix $F$, $\theta_{23}$ and $\phi$, in addition to $\delta$, in principle, must transform, as indicated in the explicit form of the SF equation~\eqref{SF-eq-DMP}: 
\begin{eqnarray} 
&& 
\left[
\begin{array}{ccc}
e^{ i \tau } & 0 & 0 \\
0 & e^{ i \sigma } & 0 \\
0 & 0 & 1 \\
\end{array}
\right]
\left[
\begin{array}{c}
\nu_{e} \\
\nu_{\mu} \\
\nu_{\tau} \\
\end{array}
\right] 
\nonumber \\
&=& 
\left[
\begin{array}{ccc}
1 & 0 &  0  \\
0 & c_{23} & s_{23} e^{ i \sigma } \\
0 & - s_{23} e^{ - i \sigma } & c_{23} \\
\end{array}
\right] 
\left[
\begin{array}{ccc}
c_{\phi}  & 0 & s_{\phi} e^{ i \tau } \\
0 & 1 & 0 \\
- s_{\phi} e^{ - i \tau } & 0 & c_{\phi} \\
\end{array}
\right] 
\left[
\begin{array}{ccc}
c_{\psi} & s_{\psi} e^{ i ( \delta + \tau - \sigma ) } &  0  \\
- s_{\psi} e^{ - i ( \delta + \tau - \sigma ) } & c_{\psi} & 0 \\
0 & 0 & 1 \\
\end{array}
\right] 
\left[
\begin{array}{ccc}
e^{ i \tau } & 0 & 0 \\
0 & e^{ i \sigma } & 0 \\
0 & 0 & 1 \\
\end{array}
\right] 
G^{\dagger} 
\nonumber \\
&\times& 
G 
\biggl\{
1 + 
\epsilon c_{12} s_{12} \sin ( \phi - \theta_{13} ) 
\mathcal{W} ( \psi, \delta; \lambda_{1}, \lambda_{2} ) 
\biggr\} 
G^{\dagger}
G 
\left[
\begin{array}{c}
\nu_{1} \\
\nu_{2} \\
\nu_{3} \\
\end{array}
\right].
\label{SF-eq-DMP2}
\end{eqnarray}
Notice that rephasing not only of the flavor states $\nu_{\alpha}$ ($\alpha=e, \mu, \tau$) but also the matter eigenstates $\nu_{j}$ ($j=1,2,3$) does not affect the oscillation probability. 

For $s_{23}^{\prime}$ and $s_{\phi}^{\prime}$, we restrict ourselves to the simple solutions $s_{23} e^{ i \sigma } = s_{23}^{\prime}$ and $s_{\phi} e^{ i \tau } = s_{\phi}^{\prime}$. Apparently, there is no other way as far as we remain in the present formulation of the SF equation~\eqref{SF-eq-DMP2}. Once we place this restriction, we have to limit the possible solutions of $\tau$ and $\sigma$ to integer multiples of $\pi$, since otherwise we have to make the mixing angles complex. We will see in the next section~\ref{sec:how-to} that it has a tremendous consequence to restrict the solution space of the SF equation. 

Under the ansatz $s_{23} e^{ i \sigma } = s_{23}^{\prime}$ and $s_{\phi} e^{ i \tau } = s_{\phi}^{\prime}$, the SF equation~\eqref{SF-eq-DMP2} can be decomposed into the following first and the second conditions: The first condition reads 
\begin{eqnarray} 
&& 
\left[
\begin{array}{ccc}
c_{\psi} & s_{\psi} e^{ i ( \delta + \tau - \sigma ) } &  0  \\
- s_{\psi} e^{ - i ( \delta + \tau - \sigma ) } & c_{\psi} & 0 \\
0 & 0 & 1 \\
\end{array}
\right] 
\left[
\begin{array}{ccc}
e^{ i \tau } & 0 & 0 \\
0 & e^{ i \sigma } & 0 \\
0 & 0 & 1 \\
\end{array}
\right] 
G^{\dagger} 
= 
U_{12} ( \psi^{\prime}, \delta + \xi),
\label{1st-condition}
\end{eqnarray}
while the second condition takes the form 
\begin{eqnarray} 
&&
\epsilon c_{12} s_{12} \sin ( \phi - \theta_{13} )
\left[
\begin{array}{ccc}
0 & 0 & - c_\psi e^{ i \alpha } 
\frac{ \Delta m^2_{ \text{ren} } }{ \lambda_{3} - \lambda_{2} } \\
0 & 0 & - s_\psi e^{ - i ( \delta + \beta) } 
\frac{ \Delta m^2_{ \text{ren} } }{ \lambda_{3} - \lambda_{1} }  \\
c_\psi e^{ - i \alpha } 
\frac{ \Delta m^2_{ \text{ren} } }{ \lambda_{3} - \lambda_{2} } & 
s_\psi e^{ i ( \delta + \beta) } 
\frac{ \Delta m^2_{ \text{ren} } }{ \lambda_{3} - \lambda_{1} } & 0 \\
\end{array}
\right] 
\nonumber \\
&=&
\epsilon c_{12}^{\prime} s_{12}^{\prime} \sin ( \phi^{\prime} - \theta_{13}^{\prime} )
\left[
\begin{array}{ccc}
0 & 0 & - s_\psi^{\prime}
\frac{ \Delta m^2_{ \text{ren} } }{ \lambda_{3} - \lambda_{2} } \\
0 & 0 & c_\psi^{\prime} e^{ - i ( \delta + \xi) } 
\frac{ \Delta m^2_{ \text{ren} } }{ \lambda_{3} - \lambda_{1} } \\
s_\psi^{\prime}
\frac{ \Delta m^2_{ \text{ren} } }{ \lambda_{3} - \lambda_{2} } & 
- c_\psi^{\prime} e^{ i ( \delta + \xi) } 
\frac{ \Delta m^2_{ \text{ren} } }{ \lambda_{3} - \lambda_{1} } & 0 \\
\end{array}
\right].
\label{2nd-condition-full} 
\end{eqnarray}

\subsection{How to solve the DMP SF equation?} 
\label{sec:how-to}

We analyze the first condition~\eqref{1st-condition}. It is not difficult to show that it entails the conditions 
\begin{eqnarray} 
&&
c_{\psi^{\prime}} 
= - s_{\psi} e^{ - i ( \alpha - \tau ) } 
= - s_{\psi} e^{ i ( \beta + \sigma ) }, 
\hspace{8mm}
s_{\psi^{\prime}} 
= c_{\psi} e^{ i ( \beta + \tau - \xi ) } 
= c_{\psi} e^{ - i ( \alpha - \sigma - \xi ) }, 
\label{1st-condition-sol}
\end{eqnarray}
and the consistency conditions for the phases 
\begin{eqnarray} 
&&
\alpha + \beta - \tau + \sigma = 0 ~~~(\text{mod.} ~2\pi), 
\hspace{8mm}
\tau - \sigma - \xi = 0, ~\pm \pi
\label{1st-sol-consistency} 
\end{eqnarray}
where all the solutions of the phases are modulo $2\pi$. In the second equation in eq.~\eqref{1st-sol-consistency}, a classification naturally appeared: 
\begin{eqnarray} 
&&
\text{ Class I: } \tau - \sigma = \xi, 
\hspace{8mm}
\text{ Class II: } \tau - \sigma = \xi \pm \pi.
\label{Class-I-II}
\end{eqnarray}

Since we have restricted $\tau$ and $\sigma$ to integer multiples of $\pi$, eq.~\eqref{1st-condition-sol} dictates that $\alpha$, $\beta$, and $\xi$ must also be integer multiples of $\pi$. That is, all the $\xi$, $\tau$, $\sigma$, $\alpha$, and $\beta$ must be integer multiples of $\pi$. 
Then, the procedure for obtaining solutions to the SF equation is: 
\begin{itemize}

\item 
To choose an ansatz for $\xi$. In this paper we try only the limited choices, $\xi = 0, \pi$.

\item 
To choose Class I or Class II. Then, find all possible solutions for $\tau$, $\sigma$, $\alpha$, and $\beta$. 

\item 
Verify the solution against the second condition~\eqref{2nd-condition-full}. 

\end{itemize}
It is interesting to observe that the A-type and B-type pairings naturally arise from the choices $\xi = 0$ or $\pi$, which ultimately came from the first condition in the SF equation~\eqref{SF-eq-DMP2}. This and the other consequences will be explained in section~\ref{sec:eight-symmetries}. Though we do not know if the solutions are complete, we did not make any serious effort to go outside of the current framework.\footnote{
A real challenge would be to formulate the SF equation with an extended exchange symmetry which involves the three mass eigenstates, which is however far beyond the scope of this paper. }

\subsection{Symmetry IA and IB in DMP}
\label{sec:IA-IB-DMP}

Let us analyze the SF equation in eqs.~\eqref{1st-condition} and \eqref{2nd-condition-full}. We first recover the $\lambda_{1} \leftrightarrow \lambda_{2}$ exchange symmetry, the one found in the original DMP work~\cite{Denton:2016wmg}, which is denoted as ``Symmetry IA-DMP'' for the classification purpose: 
\begin{eqnarray} 
&& 
\text{Symmetry IA-DMP:}
\hspace{8mm}
\lambda_{1} \leftrightarrow \lambda_{2}, 
\hspace{8mm}
c_{\psi} \rightarrow \mp s_{\psi}, 
\hspace{8mm}
s_{\psi} \rightarrow \pm c_{\psi}. 
\label{Symmetry-IA-DMP}
\end{eqnarray}
Since no transformation of the fundamental variables is involved we can set $\tau=\sigma = 0$. Following the prescription in section~\ref{sec:how-to} we try the two cases, $\xi=0$ and $\xi=\pi$. Notice that $\xi$ is the parameter that causes possible shift of $\delta$ in the transformed system, as in eq.~\eqref{1st-condition}.

\subsubsection{Case of $\xi=0$: Class I} 

It is easy to observe, with eqs.~\eqref{1st-condition-sol} and \eqref{1st-sol-consistency}, that there are the two solutions of $\alpha$ and $\beta$, $\alpha = \beta = 0$ for the upper sign and $\alpha = \pi, \beta = - \pi$ for the lower sign in eq.~\eqref{Symmetry-IA-DMP}, respectively. The consistency with the second condition~\eqref{2nd-condition-full} can be easily checked. Therefore, it is shown that Symmetry IA-DMP in eq.~\eqref{Symmetry-IA-DMP} is the solution to the SF equation in DMP. 

\subsubsection{Case of $\xi=\pi$: Class II} 

Let us explore the case of $\xi=\pi$ to look for a new solution. If we try the cases of $\alpha = \pi, \beta = - \pi$ and $\alpha = \beta = 0$, we obtain 
$c_{\psi^{\prime}} = s_{\psi}$, $s_{\psi^{\prime}} = c_{\psi}$ 
($c_{\psi^{\prime}} = - s_{\psi}$, $s_{\psi^{\prime}} = - c_{\psi}$) 
for the former (latter) case. Then, the second condition tells us that the sign flip of $s_{12}$ is required. Thus, we have found a new symmetry ``Symmetry IB-DMP'' as a solution to the SF equation: 
\begin{eqnarray} 
&& 
\text{Symmetry IB-DMP:}
\hspace{8mm} 
\theta_{12} \rightarrow - \theta_{12}, 
\hspace{8mm} 
\delta \rightarrow \delta + \pi, 
\nonumber \\
&& 
\hspace{42mm} 
\lambda_{1} \leftrightarrow \lambda_{2}, 
\hspace{10mm}
c_{\psi} \rightarrow \pm s_{\psi}, 
\hspace{10mm}
s_{\psi} \rightarrow \pm c_{\psi}. 
\label{Symmetry-IB-DMP}
\end{eqnarray}
Thus, we have found that the pairing structure of the $1 \leftrightarrow 2$ exchange symmetry with A ($\delta$ free) and B (with $\delta$) types prevails in DMP.  

A question might be: After we find more symmetries, which will be the case, why is the paring of Symmetry IA and IB mandatory, e.g., not IIA and IB? The answer to this question will be given in a clear way in the discussion of Hamiltonian viewpoint of the symmetries in section~\ref{sec:hamiltonian-view}.

\subsection{Consistency check among the transformations}
\label{sec:consistency-check}

It is important to verify the mutual consistency between the transformations which are involved in  the symmetries. We do it here for Symmetry-IA and IB-DMP, but it should be repeated when we find another new symmetry later. 

The expressions of the eigenvalues $\lambda_{1}$ and $\lambda_{2}$ can be expressed as 
\begin{eqnarray} 
&& 
\lambda_{1} \equiv 
c^2_\psi \lambda_{-} + s^2_\psi \lambda_{0} - 2 c_\psi s_\psi \mathcal{A}, 
\nonumber \\ 
&& 
\lambda_{2} \equiv 
s^2_\psi \lambda_{-} + c^2_\psi \lambda_{0} + 2 c_\psi s_\psi \mathcal{A}, 
\label{lambda-1-2}
\end{eqnarray}
where $\mathcal{A} \equiv \epsilon c_{12} s_{12} c_{\phi - \theta_{13}} \Delta m^2_{ \text{ren} }$.\footnote{
$\lambda_{-}$ and $\lambda_{0}$ are the pre-$\theta_{12}$ rotation eigenvalues, which are identical to the ones in ref.~\cite{Minakata:2015gra}. $\lambda_{-}$ and $\lambda_{0}$ are the second-largest and smallest eigenvalues in the asymptotic region $a \rightarrow + \infty$ in the normal mass ordering. }
In Symmetry-IA-DMP, the transformation between $c_{\psi}$ and $s_{\psi}$ is such that $\mathcal{A}$ remains invariant. But since $c_{\psi} s_{\psi} \rightarrow - c_{\psi} s_{\psi}$, the eigenvalue exchange $\lambda_{1} \leftrightarrow \lambda_{2}$ occurs by the $\psi$ transformation. 
In Symmetry-IB-DMP, on the other hand, the transformation between $c_{\psi}$ and $s_{\psi}$ is such that $\mathcal{A}$ flips sign while $c_{\psi} s_{\psi}$ is invariant. Therefore, the transformations of $\lambda_{i}$ and $\psi$ are mutually consistent in the both symmetries Symmetry-IA and IB in DMP. 

\subsection{Symmetry IVB-DMP}
\label{sec:IVB-DMP}

It would be a little cumbersome for the readers to follow the similar discussions for the remaining six symmetries. But we need to discuss at least one case in which the rephasing matrix $F$ matrix plays a role. Therefore, we investigate here the most involved case in which all the fundamental and dynamical parameters transform, which will be termed as ``Symmetry IVB-DMP''. The results of the remaining five symmetries will be given in the summary Table~\ref{tab:DMP-symmetry} in section~\ref{sec:eight-symmetries}, which also includes the ones discussed in the text.   

We examine the case of Class II and $\xi=\pi$. In Class II, $\tau - \sigma = \xi \pm \pi = 0$~(mod. $2\pi$), there are only the two cases, $\tau= \sigma=0$ and $\tau= \sigma=\pi$. Since the former is already examined in section~\ref{sec:IA-IB-DMP} which resulted in Symmetry IB-DMP, we concentrate here on the case $\tau= \sigma=\pi$. 

Within these restrictions, we can find the following two solutions: First solution: $\alpha=0, \beta = 0$, and Second solution: $\alpha=\pi, \beta = - \pi$. It is easy to recognize, by following the procedure described in section~\ref{sec:how-to}, that the above two solutions lead to the solutions of the first condition~\eqref{1st-condition-sol} as 
\begin{eqnarray} 
&&
c_{\psi^{\prime}} = \pm s_{\psi}, 
\hspace{8mm} 
s_{\psi^{\prime}} = \pm c_{\psi},
\label{sol-IVB-psi}
\end{eqnarray}
where the upper (lower) sign in eq.~\eqref{sol-IVB-psi} corresponds to the First (Second) solution of $\alpha, \beta$. 
Notice that due to the non-vanishing flavor basis neutrino rephasing phases, $\tau$ and $\sigma$, the vacuum mixing angles $\theta_{23}$ and $\theta_{13}$ must transform as $\theta_{23} \rightarrow - \theta_{23}$ and $\theta_{13} \rightarrow - \theta_{13}$. Then, $\theta_{13}$ transformation induces the $\phi \rightarrow - \phi$ transformation, because $\sin 2\phi \propto \sin 2\theta_{13}$~\cite{Denton:2016wmg}. Finally, the consistency with the second condition~\eqref{2nd-condition-full} requires $\theta_{12}$ transform as $\theta_{12} \rightarrow - \theta_{12}$.

Thus, we have obtained a new symmetry which we call ``Symmetry IVB-DMP'': 
\begin{eqnarray} 
&&
\text{Symmetry IVB-DMP:}
\nonumber \\
&&
\theta_{23} \rightarrow - \theta_{23}, 
\hspace{8mm}
\theta_{13} \rightarrow - \theta_{13},
\hspace{8mm}
\theta_{12} \rightarrow- \theta_{12}, 
\hspace{8mm}
\delta \rightarrow \delta + \pi, 
\nonumber \\
&&
\lambda_{1} \leftrightarrow \lambda_{2}, 
\hspace{13mm}
\phi \rightarrow - \phi, 
\hspace{13mm}
c_{\psi} \rightarrow \pm s_{\psi}, 
\hspace{10mm}
s_{\psi} \rightarrow \pm c_{\psi}.
\label{Symmetry-IVB-DMP}
\end{eqnarray}

Consistency between the $\lambda_{1} \leftrightarrow \lambda_{2}$ exchange and the $\psi$ transformation is maintained. As in the case of Symmetry-IB-DMP, $\mathcal{A} \equiv \epsilon c_{12} s_{12} c_{\phi - \theta_{13}} \Delta m^2_{ \text{ren} }$ flips sign while $c_{\psi} s_{\psi}$ is invariant. Thus, transformations of $\psi$ are consistent with the eigenvalue exchange. See eq.~\eqref{lambda-1-2}. With regard to $\psi$, $s_\psi c_\psi = \frac{1}{2} \sin 2\psi$ transform as~\cite{Denton:2016wmg} 
\begin{eqnarray} 
&& 
s_\psi c_\psi = \frac{ \epsilon c_{12} s_{12} c_{\phi - \theta_{13}} \Delta m^2_{ \text{ren} } }{ \lambda_2 - \lambda_1} 
\rightarrow s_\psi c_\psi, 
\label{sin2psi}
\end{eqnarray}
under the transformation in eq.~\eqref{Symmetry-IVB-DMP}, and $\cos 2\psi$ as 
\begin{eqnarray} 
&& 
c^2_\psi-s^2_\psi=\frac{\lambda_0 - \lambda_-}{ \lambda_2 - \lambda_1} 
\rightarrow - ( c^2_\psi-s^2_\psi ). 
\label{cos2psi}
\end{eqnarray}
Therefore, the transformations of the vacuum variables, the fundamental parameters, and the dynamical variables are mutually fully consistent. 

\subsection{The whole structure of $1 \leftrightarrow 2$ exchange symmetry in DMP}
\label{sec:eight-symmetries} 

We are ready to present all the solutions of the SF equation, thereby displaying the whole structure of $1 \leftrightarrow 2$ exchange symmetry in DMP. We do this by presenting Table~\ref{tab:DMP-symmetry} for the eight, or the pair-doubled four symmetries in DMP. Interested readers can easily follow the procedure described in section~\ref{sec:how-to} to reproduce these results. To help this task, we provide the supplementary information in Table~\ref{tab:SF-solutions} to show how each symmetry corresponds to which solution of the SF equation. As noticed before, Symmetry IA-DMP in the first line in Table~\ref{tab:DMP-symmetry} and~\ref{tab:SF-solutions} was noticed in ref.~\cite{Denton:2016wmg}, but the remaining seven symmetries are all new.  


\begin{table}[h!]
\vglue -0.2cm
\begin{center}
\caption{ All the symmetries of the $1 \leftrightarrow 2$ eigenstate exchange type in the DMP system that have been uncovered in this paper are summarized. They are found  in a systematic way by solving the SF equation with appropriate ansatz, as some of them (Symmetry IA, IB, and IVB) are fully explained in section~\ref{sec:symmetry-finder-DMP}. The symmetry denoted as e.g., ``Symmetry X'' in this Table is called as ``Symmetry X-DMP'' in the text, where X = IA, IB, IIA, IIB, IIIA, IIIB, IVA, and IVB. 
In this table the notations are such that $\lambda_{j}$ ($j=1,2$) are the eigenvalues of $2E H$, $\psi$ and $\phi$ denote $\theta_{12}$ and $\theta_{13}$ in matter, respectively.
}
\label{tab:DMP-symmetry}
\vglue 0.2cm
\begin{tabular}{c|c|c}
\hline 
Symmetry & 
Vacuum parameter transformations & 
Matter parameter transformations
\\
\hline 
\hline 
Symmetry IA & 
none & 
$\lambda_{1} \leftrightarrow \lambda_{2}$, 
$c_{\psi} \rightarrow \mp s_{\psi}$, 
$s_{\psi} \rightarrow \pm c_{\psi}$. \\
\hline 
Symmetry IB & 
$\theta_{12} \rightarrow - \theta_{12}$, 
$\delta \rightarrow \delta + \pi$. & 
$\lambda_{1} \leftrightarrow \lambda_{2}$, 
$c_{\psi} \rightarrow \pm s_{\psi}$, 
$s_{\psi} \rightarrow \pm c_{\psi}$. \\
\hline
Symmetry IIA & 
$\theta_{23} \rightarrow - \theta_{23}$, 
$\theta_{12} \rightarrow - \theta_{12}$. & 
$\lambda_{1} \leftrightarrow \lambda_{2}$, 
$c_{\psi} \rightarrow \pm s_{\psi}$, 
$s_{\psi} \rightarrow \pm c_{\psi}$. \\
\hline 
Symmetry IIB & 
$\theta_{23} \rightarrow - \theta_{23}$, 
$\delta \rightarrow \delta + \pi$. & 
$\lambda_{1} \leftrightarrow \lambda_{2}$, 
$c_{\psi} \rightarrow \mp s_{\psi}$, 
$s_{\psi} \rightarrow \pm c_{\psi}$. \\
\hline 
Symmetry IIIA & 
$\theta_{13} \rightarrow - \theta_{13}$, 
$\theta_{12} \rightarrow - \theta_{12}$. & 
$\lambda_{1} \leftrightarrow \lambda_{2}$, 
$\phi \rightarrow - \phi$, \\ 
 & & 
$c_{\psi} \rightarrow \pm s_{\psi}$, 
$s_{\psi} \rightarrow \pm c_{\psi}$ \\
\hline 
Symmetry IIIB & 
$\theta_{13} \rightarrow - \theta_{13}$, 
$\delta \rightarrow \delta + \pi$. & 
$\lambda_{1} \leftrightarrow \lambda_{2}$, 
$\phi \rightarrow - \phi$, \\ 
 & & 
$c_{\psi} \rightarrow \mp s_{\psi}$, 
$s_{\psi} \rightarrow \pm c_{\psi}$. \\
\hline 
Symmetry IVA & 
$\theta_{23} \rightarrow - \theta_{23}$, 
$\theta_{13} \rightarrow - \theta_{13}$. & 
$\lambda_{1} \leftrightarrow \lambda_{2}$, 
$\phi \rightarrow - \phi$, \\ 
 & & 
$c_{\psi} \rightarrow \mp s_{\psi}$, 
$s_{\psi} \rightarrow \pm c_{\psi}$. \\
\hline 
Symmetry IVB & 
$\theta_{23} \rightarrow - \theta_{23}$, 
$\theta_{13} \rightarrow - \theta_{13}$, & 
$\lambda_{1} \leftrightarrow \lambda_{2}$, 
$\phi \rightarrow - \phi$, \\ 
 &
$\theta_{12} \rightarrow - \theta_{12}$, $\delta \rightarrow \delta + \pi$. 
 &
$c_{\psi} \rightarrow \pm s_{\psi}$, $s_{\psi} \rightarrow \pm c_{\psi}$. \\
\hline 
\end{tabular}
\end{center}
\vglue -0.4cm 
\end{table}

In Table~\ref{tab:DMP-symmetry}, the pairing of the type-A (no $\delta$ is involved) and type-B ($\delta$ is involved) symmetries is clearly visible for all the symmetries I, II, III, and IV. It arises because, roughly speaking, the transformations $\theta_{12} \rightarrow - \theta_{12}$ and $\delta \rightarrow \delta + \pi$ are equivalent for the symmetry purposes in one case, and simultaneous usage of the both transformations produces a null transformation in the other, as their effects are both multiplicative. This structure is already implicit in the SOL convention of the $U$ matrix in which the factor $e^{ \pm i \delta}$ is attached to $s_{12}$.\footnote{
One should not misunderstand that the type-A and -B pairing of the symmetry is due to a trivial consequence of the $s_{12} e^{ \pm i \delta}$ structure. When the type-A symmetry becomes type-B,  the $\psi$ transformation changes from $\sin 2\psi \rightarrow - \sin 2\psi$ to $\sin 2\psi \rightarrow \sin 2\psi$ (Symmetry I, IV ) or vice versa (Symmetry II, III). Thus, understanding the whole structure requires the SF equation.
}
But, again, we emphasize that the symmetries are valid in the probabilities computed by using the $U$ matrix e.g., in the PDG convention. 

Can we claim that the symmetries of the $1 \leftrightarrow 2$ state exchange type discussed above are all the possible symmetries in DMP? Of course, we cannot because the ansatz we used to solve the SF equation are of limited ones. Nevertheless, in view of the remarkable completeness of the symmetry structure including the pairing of A- and B-type symmetries, the possibility may not be immediately excluded that the symmetries listed in Table~\ref{tab:DMP-symmetry} and~\ref{tab:SF-solutions} exhaust all the possible $1 \leftrightarrow 2$ state exchange symmetries in DMP. 


\begin{table}[h!]
\vglue -0.2cm
\begin{center}
\caption{ 
The case of $\tau = \sigma + \xi$ is called as Class I, and $\tau = \sigma + \xi \pm \pi$ as Class II. The labels ``upper'' and ``lower'' imply the upper and lower sign in the corresponding columns in Table~\ref{tab:DMP-symmetry}. 
}
\label{tab:SF-solutions}
\vglue 0.2cm
\begin{tabular}{c|c|c}
\hline 
Symmetry & 
Class I or II, $\tau, \sigma, \xi$ & 
$\alpha, \beta$
\\
\hline 
\hline 
Symmetry IA & 
Class I, $\tau = \sigma = 0$, $\xi = 0$ & 
$\alpha = \beta = 0$ (upper) \\ 
& &
$\alpha = \pi, \beta = - \pi$ (lower)  \\
\hline
Symmetry IB & 
Class II, $\tau = \sigma = 0$, $\xi = \pi$ & 
$\alpha = \pi, \beta = - \pi$ (upper) \\
& & $\alpha = \beta = 0$ (lower) \\
\hline 
Symmetry IIA & 
Class II, $\tau = 0, \sigma = - \pi$, $\xi = 0$ & 
$\alpha = \pi, \beta = 0$ (upper) \\
& & $\alpha = 0, \beta = \pi$ (lower)  \\
\hline 
Symmetry IIB & 
Class I, $\tau = 0, \sigma = - \pi$, $\xi = \pi$ & 
$\alpha = 0, \beta = \pi$ (upper) \\ 
& & $\alpha = \pi, \beta = 0$ (lower) \\
\hline 
Symmetry IIIA & 
Class II, $\tau = \pi, \sigma = 0$, $\xi = 0$ & 
$\alpha = 0, \beta = \pi$ (upper) \\ 
& & $\alpha = \pi, \beta = 0$ (lower) \\
\hline 
Symmetry IIIB & 
Class I, $\tau = \pi, \sigma = 0$, $\xi = \pi$ & 
$\alpha = \pi, \beta = 0$ (upper)  \\ 
 & & 
$\alpha = 0, \beta = \pi$ (lower)   \\
\hline 
Symmetry IVA & 
Class I, $\tau = \sigma = \pi$, $\xi = 0$ & 
$\alpha = \pi, \beta = - \pi$ (upper) \\ 
& &
$\alpha = \beta = 0$ (lower)  \\
\hline 
Symmetry IVB & 
Class II, $\tau = \sigma = \pi$, $\xi = \pi$ & 
$\alpha = \beta = 0$ (upper) \\ 
& &
$\alpha = \pi, \beta = - \pi$ (lower)  \\
\hline 
\end{tabular}
\end{center}
\vglue -0.4cm 
\end{table}

\section{Hamiltonian view of the $1 \leftrightarrow 2$ exchange symmetry in matter} 
\label{sec:hamiltonian-view} 

We now discuss the Hamiltonian view of the reparametrization symmetry of the $1 \leftrightarrow 2$ state exchange type in matter. We primarily treat in this section the DMP system, because the symmetries in the ZS system and in vacuum can be easily understood as one of the DMP's symmetry with appropriate replacements of the variables. It is also worth to mention that our focus on these $1 \leftrightarrow 2$ exchange symmetries in DMP is due to their rich varieties, exhibiting interesting features. 

\subsection{Hamiltonian view of the DMP system}
\label{sec:H-view-DMP}

The flavor basis Hamiltonian can be written by the two ways, by using the fundamental variables, which we call $H_{ \text{LHS} }$, or by the dynamical (i.e., Hamiltonian diagonalizing) variables, which we call $H_{ \text{RHS} }$. Of course they must be equal, 
\begin{eqnarray}
&&
H_{ \text{LHS} } = H_{ \text{RHS} }.
\label{H-vac-matter}
\end{eqnarray}
The explicit expressions of $H_{ \text{LHS} }$, and $H_{ \text{RHS} }$ in the DMP decomposed form into the unperturbed and perturbed terms read in the SOL convention as 
\begin{eqnarray}
&&
2E H_{ \text{LHS} } 
\nonumber \\
&=&
U_{23} (\theta_{23}) U_{13} (\theta_{13}) U_{12} (\theta_{12}, \delta) 
\left[
\begin{array}{ccc}
m^2_{1} & 0 & 0 \\
0 & m^2_{2} & 0 \\
0 & 0 & m^2_{3} \\
\end{array}
\right] 
U_{12} (\theta_{12}, \delta)^{\dagger} U_{13} (\theta_{13})^{\dagger} U_{23} (\theta_{23})^{\dagger}  
+ \left[
\begin{array}{ccc}
a(x) & 0 & 0 \\
0 & 0 & 0 \\
0 & 0 & 0
\end{array}
\right], 
\nonumber \\
\label{H-LHS}
\end{eqnarray}
\begin{eqnarray}
&&
2E H_{ \text{RHS} } 
\nonumber \\
&=& 
U_{23} (\theta_{23} ) U_{13} (\phi) U_{12} (\psi, \delta) 
\left\{
\left[
\begin{array}{ccc}
\lambda_{1} & 0 & 0 \\
0 & \lambda_{2} & 0 \\
0 & 0 & \lambda_{3} \\
\end{array}
\right] 
+ 
\epsilon c_{12} s_{12} \sin (\phi - \theta_{13}) \Delta m^2_{ \text{ren} } 
\left[
\begin{array}{ccc}
0 & 0 & - s_{\psi} \\
0 & 0 & c_{\psi} e^{ - i \delta} \\
- s_{\psi} & c_{\psi} e^{ i \delta} & 0 \\
\end{array}
\right]
\right\} 
\nonumber \\
&\times&
U^{\dagger}_{12} (\psi, \delta) U^{\dagger}_{13} (\phi) U^{\dagger}_{23} (\theta_{23}).
\label{H-RHS} 
\end{eqnarray}
Notice that $H_{ \text{RHS} }$ is shown as the decomposed form but it is exact, namely, no approximation is made to derive eq.~\eqref{H-RHS}. 
We denote the first and second terms in $H_{ \text{LHS} }$ in eq.~\eqref{H-LHS} as $H_{ \text{vac} }$ and $H_{ \text{matt} }$, respectively. Similarly, we denote the first and second terms in $H_{ \text{RHS} }$ in eq.~\eqref{H-RHS} as $H_0$ and $H_{1}$, respectively. That is, $H_{ \text{LHS} } = H_{ \text{vac} } + H_{ \text{matt} }$ and $H_{ \text{RHS} } = H_0 + H_{1}$. 

\subsection{Transformation of $H_{ \text{LHS} }$} 
\label{sec:H-LHS-transf}

Let us first understand how $H_{ \text{LHS} }$ in eq.~\eqref{H-LHS} transform. $H_{ \text{vac} }$ transforms only by the fundamental (i.e., vacuum mixing) parameters, and $H_{ \text{matt} }$ remains intact. For Symmetry IA, obviously $H_{ \text{vac} }$ does not transform, as no vacuum parameter transforms. For Symmetry IB, the vacuum parameters change as $\theta_{12} \rightarrow - \theta_{12}$ and $\delta \rightarrow \delta + \pi$, but they keep $U_{\text{\tiny SOL}}$ and hence $H_{ \text{vac} }$ invariant, as they come-in in the combination $s_{12} e^{ \pm i \delta}$. Therefore, $H_{ \text{LHS} }$ is invariant under the both Symmetry IA and IB. 

For Symmetry X, for X=IIA and IIB (see Table~\ref{tab:DMP-symmetry}), one can easily show that $H_{ \text{vac} }$ transforms 
the vacuum parameters transformation as 
\begin{eqnarray}
&&
H_{ \text{vac} } 
\equiv 
\left[
\begin{array}{ccc}
h_{ee} & h_{e \mu} & h_{e \tau} \\
h_{\mu e} & h_{\mu \mu} & h_{\mu \tau} \\
h_{\tau e} & h_{\tau \mu} & h_{\tau \tau}
\end{array}
\right]
\rightarrow 
\left[
\begin{array}{ccc}
h_{ee} & - h_{e \mu} & h_{e \tau} \\
- h_{\mu e} & h_{\mu \mu} & - h_{\mu \tau} \\
h_{\tau e} & - h_{\tau \mu} & h_{\tau \tau}
\end{array}
\right], 
\label{H-vac-IIA-IIB}
\end{eqnarray}
leaving a lozenge position minus. Then, the transformation property of $H_{ \text{LHS} } = H_{ \text{vac} } + H_{ \text{matt} }$ can be expressed as a phase redefinition, which is to be applied onto the flavor and mass eigenstates 
\begin{eqnarray}
&&
H_{ \text{LHS} }  \rightarrow 
\text{Rep(II)}~
H_{ \text{LHS} } ~
\text{Rep(II)}
\hspace{10mm} 
\text{for X= IIA and IIB}, 
\label{LHS-IIA-IIB}
\end{eqnarray}
where we have introduced the rephasing matrix Rep(II) as 
\begin{eqnarray} 
&&
\text{Rep(II)} = 
\left[
\begin{array}{ccc}
- 1 & 0 & 0 \\
0 & 1 & 0 \\
0 & 0 & -1
\end{array}
\right].
\label{Rep-II}
\end{eqnarray}
That is, $H_{ \text{LHS} }$ is invariant under the transformations of Symmetry IIA and IIB up to the phase redefinition of the flavor and mass eigenstates. The similar exercise can be repeated for Symmetry X, X= IIIA - IIIB and IVA - IVB, with the results: 
\begin{eqnarray}
&&
H_{ \text{LHS} }  \rightarrow 
\text{Rep(III)}~
H_{ \text{LHS} } ~
\text{Rep(III)}
\hspace{10mm} 
\text{for X= IIIA and IIIB},
\nonumber \\
&& 
H_{ \text{LHS} }  \rightarrow 
\text{Rep(IV)}~
H_{ \text{LHS} } ~
\text{Rep(IV)}
\hspace{10mm} 
\text{for X = IVA and IVB}, 
\label{LHS-II-III}
\end{eqnarray}
where we have similarly defined Rep(III) and Rep(IV) as 
\begin{eqnarray} 
&&
\text{Rep(III)} = 
\left[
\begin{array}{ccc}
- 1 & 0 & 0 \\
0 & 1 & 0 \\
0 & 0 & 1
\end{array}
\right], 
\hspace{8mm}
\text{Rep(IV)} = 
\left[
\begin{array}{ccc}
- 1 & 0 & 0 \\
0 & -1 & 0 \\
0 & 0 & 1
\end{array}
\right].
\label{Rep-III-IV} 
\end{eqnarray}
Notice that all the rephasing matrix Rep(X) does not distinguish between the A- and B-type symmetries. 
Thus, the transformation property of $H_{ \text{LHS} }$ can be expressed as a phase redefinition, which is to be applied onto the flavor and mass eigenstates. 

\subsection{Transformation of $H_{ \text{RHS} }$} 
\label{sec:H-RHS-transf} 

We move on to the transformations of $H_{ \text{RHS} }$ under Symmetry X, where X=IA, IB, IIA, IIB, IIIA, IIIB, IVA, and IVB. To make our treatment simpler we prepare some minimal formulas here. We denote that $U_{12} (\psi, \delta) \rightarrow U_{12} (X)$ under the transformations of Symmetry X. An explicit computation leads to the expressions 
\begin{eqnarray} 
&& 
U_{12} (X) 
= \left[
\begin{array}{ccc}
\mp s_{\psi} & \pm c_{\psi} e^{ i \delta} & 0 \\
\mp c_{\psi} e^{ - i \delta} & \mp s_{\psi} & 0 \\
0 & 0 & 1
\end{array}
\right] 
\hspace{10mm}
\text{for X=IA, IVA},
\nonumber \\
&&
U_{12} (X) 
= 
\left[
\begin{array}{ccc}
\pm s_{\psi} & \pm c_{\psi} e^{ i \delta} & 0 \\
\mp c_{\psi} e^{ - i \delta} & \pm s_{\psi} & 0 \\
0 & 0 & 1
\end{array}
\right]
\hspace{10mm}
\text{for X=IIA, IIIA}, 
\label{U(X)}
\end{eqnarray}
where the $\pm$ signs are commensurate with those for the A type symmetries in Table~\ref{tab:DMP-symmetry}. $U_{12} (X)$ of the B type can be obtained from the A type one by the sign flip, $\pm \rightarrow \mp$ and $\mp \rightarrow \pm$ in eq.~\eqref{U(X)}. 
 
\subsubsection{Transformation of $H_{0}$} 
\label{sec:H0-transf}

We discuss the transformation of $H_{0}$. One can show that 
\begin{eqnarray} 
&&
U_{12} (\psi, \delta) 
\text{diag} (\lambda_{1}, \lambda_{2}, \lambda_{3})
U_{12} (\psi, \delta)^{\dagger} 
\rightarrow 
U_{12} (X) 
\text{diag} (\lambda_{2}, \lambda_{1}, \lambda_{3})
U_{12} (X)^{\dagger} 
\nonumber \\
&=& 
\text{Rep(X)}
U_{12} (\psi, \delta) 
\text{diag} (\lambda_{1}, \lambda_{2}, \lambda_{3})
U_{12} (\psi, \delta)^{\dagger} 
\text{Rep(X)}. 
\label{transf-allX}
\end{eqnarray}
It is important to know that the rephasing matrces Rep(X)'s for the transformation of $H_{ \text{RHS} }$ are the same as those of $H_{ \text{LHS} }$. For X=IA and IB, Rep(I) = 1, which means that the $U_{12} (\psi, \delta)$ part of $H_{0}$ is invariant under the transformation of X=IA and IB. For the rest of X, X=II, III, and IV, Rep(X) are given in eqs.~\eqref{Rep-II} and \eqref{Rep-III-IV} which imply that the $U_{12} (\psi, \delta)$ part of $H_{0}$ is invariant under the transformation up to the rephasing matrix Rep(X).\footnote{
As a matter of fact, Rep(IV) does nothing as far as the transformation property of $H_{0}$ is concerned for X=IVA and IVB. But, it is introduced to match the transformation property of $H_{ \text{LHS} }$.
}
Notice again that Rep(X) is the same for the A and B-types for all X. 

\subsubsection{Transformation of $H_{1}$} 
\label{sec:H1-transf}

Similarly, the transformation of the $U_{12} (X)$ part of $H_{1}$ can be worked out to lead to 
\begin{eqnarray} 
&&
U_{12} (\psi, \delta) 
\left[
\begin{array}{ccc}
0 & 0 & - s_{\psi} \\
0 & 0 & c_{\psi} e^{ - i \delta} \\
- s_{\psi} & c_{\psi} e^{ i \delta} & 0 \\
\end{array}
\right] 
U_{12} (\psi, \delta)^{\dagger} 
\nonumber \\
&\rightarrow&
U_{12} (X) 
\left[
\begin{array}{ccc}
0 & 0 & - s_{\psi} \\
0 & 0 & c_{\psi} e^{ - i \delta} \\
- s_{\psi} & c_{\psi} e^{ i \delta} & 0 \\
\end{array}
\right] \biggr |_{ \text{X transformed} }
~U_{12} (X)^{\dagger} 
\nonumber \\
&=& 
\text{Sign(X)}
\text{Rep(X)} 
U_{12} (\psi, \delta) 
\left[
\begin{array}{ccc}
0 & 0 & - s_{\psi} \\
0 & 0 & c_{\psi} e^{ - i \delta} \\
- s_{\psi} & c_{\psi} e^{ i \delta} & 0 \\
\end{array}
\right]
U_{12} (\psi, \delta)^{\dagger} 
\text{Rep(X)} 
\nonumber 
\end{eqnarray}
where the prescription $\vert _{ \text{X transformed} }$ implies that the transformed variables of $\psi$ and $\delta$ by the X transformation are to be inserted, which of course depend on X. 
Rep(X) is the same as defined in eqs.~\eqref{Rep-II} and \eqref{Rep-III-IV}. Sign(X) = + for IA, IIB, IIIA, IVB, and Sign(X) = - for IB, IIA, IIIB, IVA. 

In order for $H_{1}$ to be invariant under these transformations the minus sign from Sign(X) must be cancelled by an extra minus sign supplied from the pre-factor $\epsilon c_{12} s_{12} \sin (\phi - \theta_{13}) \Delta m^2_{ \text{ren} }$ in eq.~\eqref{H-RHS}. It is not difficult to show that it indeed occurs. For IB and IIA, flipping the sign of $\theta_{12}$ absorbs the minus sign. For IIIB and IVA flipping the sign of $\theta_{13}$-$\phi$ cancels the minus sign. For IA and IIB no sign change occurs in the pre-factor, and for IIIA and IVB the sign change occurs twice, by flipping $\theta_{12}$ and $\theta_{13}$-$\phi$, resulting no net change in sign. These features can be easily confirmed by consulting to Table~\ref{tab:DMP-symmetry}. 

So far we have discussed how the transformations affect on the part $\mathcal{O}$ sandwiched by $U_{12} (X)$ and $U_{12} (X)^{\dagger}$ in $H_{ \text{RHS} }$. The rephasing matrix Rep(X) that appeared in the transformed part as Rep(X) $U_{12} (X) \mathcal{O} U_{12} (X)^{\dagger}$ Rep(X) must be extracted to the both end of the Hamiltonian. It is good to see that passing through the rephasing matrix remedies the flipped sign of $\sin \theta_{ij}$ for $ij = 23, 13$, and $12$. Also one might wonder if passing through the rephasing matrix might affect the other rotation matrix when correcting the flipped sign of only $\sin \theta_{23}$ and $\sin \theta_{13}$ in Symmetry-II and III, respectively. Upon computation, one recognizes that this is carefully avoided. 

Thus, we have shown that $H_{ \text{RHS} }$ is invariant up to the phase redefinitions of the flavor and mass eigenstates with the same rephasing matrix Rep(X) as used for $H_{ \text{LHS} }$ for each Symmetry X. The fact that the different rephasing matrices Rep(X) are necessary for X=I, II, III, and IV, but they are the same for the type-A and type-B cases in each Symmetry X. It justifies the type-A and type-B pairing scheme given in Table~\ref{tab:DMP-symmetry}. 

Finally, we note that essentially the same manipulation as used to show the invariance of $H_{0}$ in Symmetry IA and IB can be used to show that the Hamiltonian of the ZS system as well as in vacuum is invariant under Symmetry IA and IB.

\subsection{Implications of symmetry as a Hamiltonian symmetry}
\label{sec:implication}

In conclusion of our treatment in the previous sections, we have shown that the both $H_{ \text{LHS} }$ and $H_{ \text{RHS} }$ in the flavor basis Hamiltonian~\eqref{H-LHS} and~\eqref{H-RHS} are invariant under the transformations of the eight $1 \leftrightarrow 2$ exchange symmetries, Symmetry IA, IB, $\cdot \cdot \cdot$, IVB given in Table~\ref{tab:DMP-symmetry}, up to the phase redefinitions of the flavor and mass eigenstates except for IA and IB. Therefore, these symmetries are the Hamiltonian symmetries. In section~\ref{sec:nature-symmetry-matter}, we will clarify the relationship between this statement and the similar one made in ref.~\cite{Martinez-Soler:2019nhb}. 

What is the implication of the symmetry being the Hamiltonian symmetry? This is one of the most important questions in this paper, and we can make the following statements: 
\begin{itemize}

\item 
The Hamiltonian symmetry holds to all orders of perturbation theory. 

\item 
The symmetry is valid for varying density matter profile, as far as the adiabaticity condition holds, that is, under a slow variation of the matter density.\footnote{
We suspect that a Hamiltonian symmetry implies the symmetry in the oscillation probability for any matter density profile as far as it is non-singular. While we look forward investigating this point, for the time being, we try to remain in the safe side by imposing the adiabaticity condition. 
}

\end{itemize}

\section{Nature of the $1 \leftrightarrow 2$ exchange symmetry in neutrino oscillation in matter}
\label{sec:nature-symmetry-matter} 

Can we draw a unified picture of all the $1 \leftrightarrow 2$ exchange symmetries in matter we have uncovered so far? 
Symmetry IA-ZS, IB-ZS, and IA-DMP are the pure dynamical symmetries in the sense that the symmetry transformations are generated purely by the Hamiltonian-diagonalizing, or the matter-dressed variables, and no fundamental variable (i.e., variable in the original Hamiltonian~\eqref{H-LHS}) is involved. By the symmetry transformations $H_{ \text{LHS} }$ in eq.~\eqref{H-LHS} is left intact, whereas $H_{ \text{RHS} }$ transforms. But, of course, $H_{ \text{RHS} }$ remains invariant without the rephasing matrix, as Symmetry IA-ZS, IB-ZS, and IA-DMP are symmetries, as confirmed explicitly for the last one in section~\ref{sec:H-RHS-transf}. 

On the other hand, in Symmetry IIA, IIB, IIIA, IIIB, IVA, IVB in DMP, $H_{ \text{LHS} }$ transforms by a part or all of the fundamental variables transformations. Even more interestingly $H_{ \text{RHS} }$ transforms by the both fundamental as well as the dynamical variables. As can be seen most clearly in Symmetry IIA, IIB, IIIA, and IIIB in DMP, one cannot take a view that the transformations of the fundamental variables induce those of the dynamical variables. Of course the transformations of the dynamical variables never generate the transformations of the fundamental variables, as is evident in Symmetry IA-ZS, IB-ZS, and IA-DMP. Therefore, to summarize: 
\begin{itemize}

\item 
In the reparametrization symmetry with the $1 \leftrightarrow 2$ state exchange in neutrino oscillation in matter, none of the fundamental and the dynamical variables that take part in the symmetry transformations are more basic.\footnote{
This characteristic may be reminiscent of ``nuclear democracy'' or ``bootstrap'', the phrases for a feature of the $S$ matrix theory to stress that none of the elementary or composite particles is  fundamental~\cite{N-democracy}. }
They are just equally important and, of course, transform in a mutually consistent way with each other.

\end{itemize}

As mentioned in section \ref{sec:introduction}, we have stated previously~\cite{Martinez-Soler:2019nhb} about Symmetry IA-DMP that ``these symmetries as the dynamical symmetry, as opposed to the Hamiltonian symmetry''. The statement refers to the feature that $H_{ \text{LHS} }$ in eq.~\eqref{H-LHS} does not transform by the Symmetry IA. But, now under more complete view with the both $H_{ \text{LHS} }$ and $H_{ \text{RHS} }$ in our sight, we say that Symmetry IA is also the Hamiltonian symmetry, as $H_{ \text{RHS} }$ receive the transformation under Symmetry IA, albeit it is invariant, of course. 

\section{Concluding remarks} 
\label{sec:conclusion}

In this paper we have discussed the $1 \leftrightarrow 2$ state relabelling, or reparametrization symmetry in neutrino oscillation in matter. We have developed a systematic method called the ``Symmetry Finder'', for uncovering symmetry in the neutrino oscillation probability in matter. We first noticed that the Zaglauer-Schwarzer system possesses Symmetry-IA-ZS and IB-ZS (section~\ref{sec:symmetry-finder-matter}), the matter equivalent of the known symmetries in vacuum~\cite{Parke:2018shx}. Then, we have applied the Symmetry Finder (FS) scheme to the Denton {\it et al.}~\cite{Denton:2016wmg} (DMP) framework formulated to first order in perturbation theory to uncover new symmetries of the $1 \leftrightarrow 2$ state exchange type. Thanks to this powerful method of the SF equation, we have identified the eight symmetries,  Symmetry IA, IB, IIA, IIB, IIIA, IIIB, IVA, IVB in DMP, which are summarized in Table~\ref{tab:DMP-symmetry}. They are all new symmetries apart from IA, which was found in ref.~\cite{Denton:2016wmg} and has stimulated our interests toward this investigation. 

We suspect that the above eight symmetries may be a complete set of the reparametrization symmetry of $1 \leftrightarrow 2$ state exchange type, because of the special features of the above set of the symmetries,  exhausting the possible ways of usage of the vacuum parameter transformation, as well as the A-type ($\delta$ free) and B-type (with $\delta$) paired structure. But, since our ansatz used to solve the SF equation is quite limited one, the above suspect cannot be a too strong one. Notwithstanding whether the suspect is true or not, to the best of our knowledge, we believe that this is the first systematic treatment of the symmetry hunting method in the neutrino oscillation probability in matter. 

Probably, one of the nicest features in our treatment in this paper is ascribed to demonstration of all the $1 \leftrightarrow 2$ state exchange reparametrization symmetry as the Hamiltonian symmetry. That is, the symmetry transformations make the flavor basis Hamiltonian invariant, possibly up to the neutrino flavor and mass eigenstate rephasing factors. Once a symmetry is elevated to a Hamiltonian symmetry, it follows that the symmetry holds to all orders in perturbation theory, even though it was originally found by using the first-order perturbation theory. It also follows that the Hamiltonian symmetry is valid for systems with varying matter densities. 

We have restricted our discussions into the symmetry whose transformations keep the system into our world, thereby excluding the ones such as $\Delta m^2_{31} \rightarrow - \Delta m^2_{31}$, or $\theta_{12} \leq \frac{\pi}{4}$ to $\theta_{12} \geq \frac{\pi}{4}$. Then, our symmetry search has to be limited inherently to reparametrization or relabelling symmetries, apart from space-time symmetries such as CPT or CP. To our view this is an acceptable limitation. Then, one may ask what is more serious limitations that can or cannot be circumvented. 
In section~\ref{sec:ij-exchange}, we have already discussed a difficulty in formulating the SF framework to include the the $1 \leftrightarrow 3$ and $2 \leftrightarrow 3$ exchange symmetry. A real final challenge would be to formulate the SF equation with an extended exchange symmetry which involves the three mass eigenstates, which is however outside the scope of this paper. 

One may ask: What is the utility of such symmetries? Our answer would be: 
\begin{itemize}
\item 
The minimum practical utility of such symmetry is to serve for a consistency check of the calculated results of the oscillation probabilities.

\item 
Understanding nature of the symmetry might produce interesting physical interpretations.   

\end{itemize}
We note that the first merit is not so negligible if a researcher has to derive the expression of the oscillation probability in an isolated place which could happen in pandemic eras. 

However, the second point above may be more intriguing and appealing. Therefore, let us examine it a little further. An overview of the features of the symmetries discussed in this paper may be summarized as the mixed fundamental and dynamical symmetry, because the transformations are carried out partly by the fundamental (i.e.,~vacuum), and partly by the Hamiltonian-diagonalizing (matter-dressed) variables. It appears that none of them is more basic in nature for neutrino oscillations in matter, another example of ``nuclear democracy''. 

But, then one may wonder why dynamical symmetry could arise from the system under the matter potential, but no neutrino self-interactions. Let us consider a high-density neutrino gases in which neutrino-neutrino interactions play an important role. Such a fully interacting neutrino system would be the best place for investigation of dynamical symmetry in its genuine sense. 
A mean-field treatment of such a system may result in the neutrino system with the external neutrino potential in its leading-order approximation. This system may be similar to the one we have discussed in this paper, apart from difference between the external neutrino and electron potentials. If this reasoning is correct, we could have started seeing the dynamical variables and the symmetry associated with them. Hence, it would be very interesting to address the symmetry and mutual relationship between the fundamental and dynamical variables in such a system as the high density neutrino gases~\cite{Pantaleone:1992eq,Samuel:1993uw}. A particular aspect of the similar nonlinear system which are relevant for supernova neutrinos has been much discussed recently, see e.g., ref.~\cite{Hannestad:2006nj} and the references therein. 

\acknowledgments
The author thanks Peter Denton and Stephen Parke for useful discussions and communications. The Symmetry Finder equation in vacuum, eq.~\eqref{symmetry-finder-vacuum} in ref.~\cite{Parke:2018shx}, has first been brought to his attention by Stephen Parke.

\end{document}